\documentstyle[12pt]{article}
\newcommand{\ket}[1]{\vert #1 \rangle}
\newcommand{\hhconfig}{ (\pi{\rm h}_{11/2})^1(\nu{\rm h}_{11/2})^1 }
\newcommand{\ectp}{ E_{\rm c}(2^{+}) }
\newcommand{\qq}{ Q_{\rm p} \! \cdot Q_{\rm n} }
\begin{document}
\title{
Roles of triaxiality and residual interaction \\
in signature inversions of $A \sim 130$ \\
odd-odd nuclei
}
\author{
N.~Tajima \\
Institute of Physics, College of General Education, \\
University of Tokyo, Komaba, Meguro, Tokyo 153, Japan.
}
\date{\today}
\maketitle

\baselineskip=0.600cm

\begin{abstract} 
Rotational bands with $\hhconfig$ configurations are studied using a
particle-rotor model in which a proton and a neutron quasiparticles
interacting through a zero-range force are coupled with a triaxial
rotor. It is shown for $^{124}$Cs that one can reproduce the
signature dependence of energy and B(M1)/B(E2) ratio best when one
takes into account $\gamma$-deformations with irrotational-flow moment
of inertia in addition to the proton-neutron interaction proposed by
Semmes and Ragnarsson.  Including both effects,
a systematic calculation of signature
splittings is performed for Cs, La isotopes and $N$=75 isotones
to be compared with experiments.
Discussions are also done on the
deficiencies of the cranking model concerning
its applicability to signature inversion phenomena in odd-odd nuclei.
\end{abstract}

\section{Introduction} \label{s_intro}

In the cranking model, the order of single-particle energy levels can
be reversed between a favored-signature state and its
unfavored-signature partner when the axial symmetry is violated and
the shortest axis is chosen as the rotation
axis\cite{FM83,BFM84,OT88}.  This type of rotation is called a {\em
positive-$\gamma$ rotation} in so-called Lund convention\cite{ALL76}.
The {\em signature-inversion phenomenon} is expected to manifest
itself in experiment
as the reversion of the staggering of rotational spectrum
because, as discussed in
sect.~\ref{s_sign}, the signature quantum number can {\em usually}
be associated
with the parity of the total angular momentum $I$ (or $I-\frac{1}{2}$
for odd-$A$ nuclei).

Signature inversions have been found systematically in regions of mass
number $A \sim 130$ and $A \sim 160$.  They occur only in
multi-quasiparticle (qp) bands.  For 2qp bands of odd-odd nuclei,
inversions take place in low-spin regions including the bandhead
states.  For odd-$A$ nuclei, they happen at higher spins than the
first backbendings, {\it i.e.}, in 3qp bands consisting of two
rotation-aligned qp's of proton or neutron and a deformation-aligned
qp of the other type of nucleon. A brief review of the experimental
facts is given in ref.~\cite{KFH92}.

When one tries to explain these signature inversions in terms of
triaxiality, one encounters a serious problem that, assuming as usual
that the moment of inertia has the same shape dependence as that of
irrotational flow, the nucleus prefers to rotate around its
intermediate-length principal axis, not the shortest one.  This type
of rotation, which is called a {\em negative-$\gamma$ rotation},
increases the signature splitting in the normal direction rather than
reverse it.

An {\it ad hoc} prescription
to realize a positive-$\gamma$ rotation is to
exchange the components of the moment of inertia between the shortest
axis and the intermediate-length axis by hand\cite{HM83}.  In this
paper, a moment of inertia having such shape dependence is called a
{\em $\gamma$-reversed} one.

The irrotational-flow moment of inertia is much more reasonable than
the $\gamma$-reversed one.  The former is naturally derived by
separating the quadrupole oscillations around a
spherical shape between rotation and vibration\cite{BM75}.  It is also
supported by a microscopic calculation based on the cranking
formula\cite{HH89}.

Our previous calculation using a particle-triaxial-rotor model, too,
supports it\cite{Ta90}:  We have treated all the nucleons in intruder
orbitals ($\pi$h$_{11/2}$ and $\nu$i$_{13/2}$) in a microscopic way.
These nucleons give rise to as much as half of the moment of
inertia of the entire nucleus.
This particle-rotor system does not show any signature
inversions, even if the rotor is given a $\gamma$-reversed moment of
inertia,
implying that a complete shell-model
configuration-mixing calculation, if possible, will probably
be approximated better with
an irrotational-flow moment of inertia rather than a
$\gamma$-reversed one.

Therefore ingredients other than triaxiality have to be taken into
account. Considering that inversions are always found in multi-qp
bands, the residual interaction between qp's seems important.
Hamamoto has analyzed a 3qp band in $^{157}$Ho employing a
$\qq$ force as the residual interaction\cite{Ha86}.
She concluded that an interaction strength larger than the
standard one by an order of magnitude
was necessary to reproduce the signature inversion.
Ikeda and Shimano have succeeded in reproducing
signature inversions in 3qp bands using the standard $\qq$ interaction,
but they have needed to change the sign of the
rotation-$\gamma$-vibration coupling\cite{IS89}, which corresponds to
the usage of a $\gamma$-reversed moment of inertia.

Semmes and Ragnarsson have employed a zero-range residual interaction
in a model in which a proton and a neutron qp's are coupled with a
triaxial rotor and applied the model to the $(\pi{\rm
h}_{11/2})^1(\nu{\rm i}_{13/2})^1$ band in $^{152}$Eu\cite{SR90,SR92}
and the $\hhconfig$ band in $^{120}$Cs\cite{CLJ92}.  These
applications are successful in reproducing the angular
momentum ($I_{\rm inv}$) at which the signature splitting changes the
sign.

In this paper, we adopt the framework of Semmes and Ragnarsson and
intend to improve their results.  Although they concluded that the
best way is to employ the proton-neutron (pn) interaction without
$\gamma$-deformations, we expect that triaxiality must be
considerably affecting signature-dependent quantities so long
as low-energy $\gamma$-bands exist in the neighboring even-even nuclei.
We choose an odd-odd nucleus $^{124}$Cs, whose experimental
spin-assignments are very reliable\cite{KFH92}, and perform an
intensive study to see what kind of combination of
$\gamma$-deformations and pn interactions is most suitable to
reproduce signature dependence in energy and B(M1)/B(E2) ratio.  Then,
using the best combination, we make a systematic calculation of
signature splittings for Cs, La isotopes and $N$=75 isotones and
compare the results with experiment.

Signature inversion phenomena in odd-odd nuclei have also been
studied in other theoretical frameworks, e.g., the cranking
model\cite{Ma91} and the interacting-boson-fermion model\cite{Yo93}.

In sect.~\ref{s_short}, in order to stress the necessity to use the
particle-rotor model, we point out two deficiencies of the cranking
model concerning its applicability to signature-inversion phenomena.
In sect.~\ref{s_vpn}, we check the reasonableness of the strength of
the pn interaction proposed by Semmes and Ragnarsson.  The model is
explained in sect.~\ref{s_model}.  The results of calculations are
given in sects.~\ref{s_res}.  The contents of this paper are
summarized in the last section.

\section{Inadequacy of the cranking model} \label{s_short}

In this section, we point out two shortcomings of the cranking model,
which make us question the applicability of the model to signature
inversion phenomena in odd-odd nuclei.
The aim is to illustrate the necessity to invoke
a particle-rotor model rather than the cranking model, although the
former is more phenomenological and yet more laborious to solve than
the latter.

\subsection{Non-unique correspondence
between signature and the parity of $I$}
\label{s_sign}

First, we discuss the relation between the parity of the total angular
momentum $I$ (or $I-\frac{1}{2}$ for odd-$A$ systems) and the
signature quantum number.
The former is what is observed in experiments, while the latter is
peculiar to the cranking model, in which the nucleus is assumed to
perform a uniform rotation around a space-fixed axis.  The signature
quantum number is defined as the eigenvalue of an operator,
\begin{equation} \label{e_rx}
      \hat{\cal R}_x \equiv \exp ( -i \pi {\hat{I}}_x ),
\end{equation}
which rotates the wavefunction of the system by $180^{\circ}$ around
the $x$-axis.  The $x$-axis is taken to be parallel to the cranking
axis, which coincides with one of the principal axes of the quadrupole
deformation.  ${\cal R}_x$ is a good quantum number in the cranking
model because both of the Coriolis field
($-\omega_{\rm rot}{\hat{I}}_{x}$)
and the Nilsson potential commute with ${\hat{\cal R}}_x$.
Now, if $I_x=I$ for the states (of the real nucleus, not of
the cranking model) under consideration, the signature and the parity
of $I$ obviously have one-to-one correspondence.  There seems to be no
justifications for identifying the two quantities which does not
depend on the assumption of $I_x=I$.

The condition $I_x=I$ is not fulfilled, however, when the fluctuation
$\langle \hat{I}_y^2 + \hat{I}_z^2 \rangle$
is not negligible, say larger than $\sim 2I$.
($I_y^2+I_z^2$ takes on $I$ for $I_x=I$ and $3I-1$ for $I_x=I-1$.
At the middle point, $I_y^2+I_z^2=2I-\frac{1}{2}$.)
For example, it is not negligible when quasiparticles are excited in
orbitals having large $\Omega$ quantum number and/or when a triaxially
deformed nucleus does three-dimensional rotations
\footnote{
A familiar example is the $\gamma$-band of the triaxial
rotor\cite{DF58}.  Although all of the three signatures $r_1$, $r_2$,
and $r_3$ are taken to be $+1$\cite{BM75}, the band includes odd-spin
members.
}.
One should be very careful in applying the cranking model to these cases.

The deviation of $I_x$ from $I$ is larger at smaller spins,
while signature inversions in odd-odd nuclei take place at low spins.
This is one of the reasons why we employ the particle-rotor model,
to which $I_x \not= I$ does not matter.

When we explain the results of calculations with particle-rotor model,
we use the word {\em signature} to signify the parity of $I$.
For example, concerning $\hhconfig$ configurations, states with
odd (even) $I$ are called favored (unfavored) signature states.

\subsection{Mixing of the single-particle signature} \label{s_quart}

Second, we examine the ability of the cranking model to treat 2qp
configurations.  In the cranking model, it is assumed that each
particle has a definite signature.  We show in this subsection,
however, that the signature of each particle can be mixed completely,
even when the total signature remains a rather good quantum number.

We consider only pure-$j$ orbitals. For our purpose, it is necessary
to distinguish between rotation-aligned (RA) orbitals and
deformation-aligned (DA) ones because they play different roles.

RA orbitals are labeled by the $x$-component $m_x$ of the angular
momentum and are signature eigenstates:
\begin{equation}
      \hat{\cal R}_x|j,m_x\rangle=e^{-i \pi m_x}|j,m_x\rangle.
\end{equation}
Orbitals
$\lbrace \ket{j,j}, \ket{j,j-2}, \cdots , \ket{j,-j+1} \rbrace$
have a signature $e^{-i \pi j}$, while orbitals
$\lbrace \ket{j,j-1}, \ket{j,j-3}, \cdots , \ket{j,-j} \rbrace$
have the opposite signature $e^{i \pi j}$.
The former set of orbitals is called the {\em favored} signature
states because it includes the orbital $\ket{j,j}$, which has the
smallest expectation value of the Coriolis field ($-\omega_{\rm rot}
\hat{j_x}$).  The latter set is called the {\em unfavored} signature
states.

For DA orbitals, the signature eigenstates are obtained by taking
linear combinations of a pair of degenerated Nilsson orbitals, whose
$z$-component of the angular momentum $\Omega$ is a good quantum number,
\begin{equation} \label{e_da}
      \Bigl\vert j,| \Omega |,{\cal R}_x=e^{\pm i \pi j} \Bigr\rangle =
      2^{-1/2} (\ket{j,\Omega} \mp \ket{j,- \Omega}).
\end{equation}
It can be shown in the lowest order perturbation theory that
orbitals with ${\cal R}_x$ = $e^{- i \pi j}$ ($e^{i \pi j}$) are favored
(unfavored) by the Coriolis field, in agreement with the
favored (unfavored) signature for RA orbitals.

We consider 2qp configurations of $(j_{\rm p})^1 (j_{\rm n})^1$ type
consisting of a proton and a neutron.
For each qp of kind $\tau$ (=p or n), we take into account a favored
orbital $f_{\tau}$ and an unfavored one $u_{\tau}$.  If they are RA
orbitals, we take
$\ket{f_{\tau}}=\ket{j_{\tau},m_x=j_{\tau}  }$ and
$\ket{u_{\tau}}=\ket{j_{\tau},m_x=j_{\tau}-1}$.
In case of DA orbitals, those of
eq.(\ref{e_da}) are used.  There are four intrinsic states
$
      \lbrace f_{\rm p} f_{\rm n}, u_{\rm p} u_{\rm n},
      f_{\rm p} u_{\rm n}, u_{\rm p} f_{\rm n} \rbrace
$
(the signature quartette).

In fig.~\ref{f_q}, we show how the quartette is grouped into
rotational bands.
It is supposed that $j_{\rm p}-j_{\rm n}$ is even.
The first portion shows a case in which two qp's are in RA
orbitals.
Because collective transitions of the rotor do not occur
between states with different RA orbitals,
there exist four
bands based on each member of the quartette.

The second portion treats a case where the proton (neutron) qp is in
RA (DA) orbitals.  The third portion is obtained by exchanging the
types (RA or DA) of orbitals between proton and neutron.  In both
cases, the difference of the signature of the RA orbital makes
different bands, while that of the DA orbital gives rise to the
odd-even staggering within each band.

The independent-qp picture has been applicable so far.  It no longer
holds when both qp's are in DA orbitals, as shown in the last portion.
The signature quartette should be mixed up to construct states labeled
with $K$ = $| \Omega_{\rm p}$ + $\Omega_{\rm n} |$, because the
rotational energy depends on $K$ as $E_I$ $\propto$ $I(I+1)-K^2$.  The
band with
$K$ = $K_{>}$ $\equiv$ $\Bigl| |\Omega_{\rm p}| + |\Omega_{\rm n}| \Bigr|$
has a wavefunction
$f_{\rm p} f_{\rm n} + u_{\rm p} u_{\rm n}$ for odd $I$
and
$f_{\rm p} u_{\rm n} + u_{\rm p} f_{\rm n}$ for even $I$, while
the band with
$K$ = $K_{<}$ $\equiv$ $\Bigl| |\Omega_{\rm p}| - |\Omega_{\rm n}| \Bigr|$
has
$f_{\rm p} f_{\rm n} - u_{\rm p} u_{\rm n}$ for odd $I$
and
$u_{\rm p} f_{\rm n} - f_{\rm p} u_{\rm n}$ for even $I$
(not normalized).
Therefore the cranking model cannot treat $K$-bands with multi-qp
intrinsic states
\footnote{
Although the tilted-axis cranking model\cite{Fr93} can produce
multi-qp $K$-band states, it does not conserve ``the signature quantum
number'' for the $\pi$-rotation around the cranking axis,
not to mention of the parity of $I$.
}.

In practice,
ideal ({\it i.e.,} pure-$m_x$) RA orbitals hardly exist and the
wavefunctions of 2qp bands inevitably contain the components of
$K$-bands.  In this respect, too, it is worth while using the
particle-rotor model for odd-odd nuclei.

Let us discuss briefly about the signature quartette.  In all the
portions of the figure except the first one, the phase of the
odd-even staggering of the higher-lying band is opposite to that of
the lower-lying one.  This feature may be useful to tell which pair of
bands makes the signature quartette.
The $\hhconfig$ bands in Cs isotopes
are roughly classified to the second case
from the fermi levels.
For $^{124}$Cs, our calculation predicts that
the first excited band of this configuration is
at about 0.6~MeV from the yrast band.
The odd-even staggerings of the lowest two bands are of opposite phase
for spins both below and above $I_{\rm inv}$.
They seem to make a signature quartette.

\section{The proton-neutron interaction} \label{s_vpn}

In refs.~\cite{SR90,SR92,CLJ92,BP76}, a zero-range interaction of the
following form
\footnote{
In refs.~\cite{SR90,SR92,BP76}, a different factor
     $(\pi b^3 / 2^{1/2})^{1/2}$
is printed instead of
     $(\pi /2)^{1/2} b^3$
of eq.~(\ref{e_vpn}).  However, eq.~(\ref{e_vpn}) is the correct
definition in the sense that it is consistent with the
particle-rotor-model code of Semmes and Ragnarsson and that the code
reproduces\cite{Se93} the matrix elements tabulated in
ref.~\cite{BP76}.  Therefore all the numerical results in
refs.~\cite{SR90,SR92,CLJ92,BP76} are completely correct if the
definition is changed to eq.(\ref{e_vpn}).
}
has been used as the residual force between the unpaired proton and
neutron in odd-odd nuclei.
\begin{equation} \label{e_vpn}
      V_{\rm pn}=4 \pi \sqrt{\frac{\pi}{2}} b^3
      \delta(\mbox{\boldmath $r$}_{\rm p}
            -\mbox{\boldmath $r$}_{\rm n})
      (u_0 + u_1 \mbox{\boldmath $\sigma$}_{\rm p} \! \cdot
                 \mbox{\boldmath $\sigma$}_{\rm n}),
\end{equation}
where $b$ $\equiv$ $(\hbar / m \omega)^{1/2}$.
Boisson and Piepenbring have obtained $u_1$ = $-0.8$~MeV\cite{BP76}
through least-squares fits to the GM shifts\cite{GM58},
leaving $u_0$ undetermined.
Semmes and Ragnarsson
have proposed a strength $u_0=-7.2$~MeV\cite{SR90} by
determining the ratio $u_0/u_1$ from $V_T$/$V_S$$\sim$0.6, the
experimental tendency of the ratio of the spin-triplet energy to the
singlet one\cite{dF74}.  They have found that this $u_0$ leads to
accurate reproductions of $I_{\rm inv}$
for $^{120}$Cs\cite{CLJ92} and $^{152}$Eu\cite{SR90,SR92}.

\subsection{How strong is the $V_{\rm pn}$ of Semmes and Ragnarsson ?}
\label{s_str}

In ref.\cite{Ha86}, it has been concluded that a $\qq$ interaction
more than ten times as strong as the standard one is necessary in
order to reproduce a signature inversion in a
     $(\pi{\rm h}_{11/2})^1(\nu{\rm i}_{13/2})^2$
band.  It should be examined whether $V_{\rm pn}$ (eq.~(\ref{e_vpn}))
is so unreasonably strong or not.

By expressing the angular part of the delta function with spherical
harmonics, $V_{\rm pn}$ can be expanded as
\begin{equation} \label{e_yy}
      V_{\rm pn} =
      V_0 \sum_{l} Y_l (\hat{r}_{\rm p}) \cdot Y_l (\hat{r}_{\rm n}) +
      V_1 \sum_{l} Y_l (\hat{r}_{\rm p}) \cdot Y_l (\hat{r}_{\rm n}) \;
      \mbox{\boldmath $\sigma$}_{\rm p} \! \cdot
      \mbox{\boldmath $\sigma$}_{\rm n},
\end{equation}
where $V_0$ = $-11.2$~MeV and $V_1$ = $-1.2$~MeV
(estimated for $\hhconfig$ configurations).
The transferred angular momentum $k$ between the proton and the
neutron can take on $l$ ($l$~and~$l \pm 1$) for the terms in the first
(second) summation.  As far as
     $(j_{\rm p})^1 (j_{\rm n})^1$-type
configurations concern, the two-body matrix elements of the right-hand
side of eq.~(\ref{e_yy}) vanish for terms with odd $l$ (due to parity
conservation) and for contributions with $k$=$l$ from the terms in the
second summation\cite{dT63}.  It follows that an interaction with
$k$=2 is contained only in the spin-independent part of the
interaction and its strength is $-11.2$~MeV.

An estimate of the strength of the $\qq$ force is found in
ref.\cite{BM75}, which gives,
\begin{equation} \label{e_vqq}
      V_{\rm QQ} = -7.7 \; \mbox{[MeV]} \; \times \;
      Y_2(\hat{r}_{\rm p}) \cdot Y_2(\hat{r}_{\rm n}),
\end{equation}
for $\hhconfig$ configurations in $A$=124 nuclei.
(We used eqs.(6-78,127,380) of ref.~\cite{BM75}.
The strength is
$(X_0-X_1) \langle r_{\rm p}^2 \rangle \langle r_{\rm n}^2 \rangle$.)
Hence, concerning the $k$=2 component, $V_{\rm pn}$ is stronger than
$V_{\rm QQ}$ only by a factor of $1.5$.

$V_{\rm pn}$ has, however, other multipole components.
To see their effects, we have plotted the two-body matrix elements,
\begin{equation} \label{e_gj}
      g_J = \langle \hhconfig ; JM | V_{\rm pn} |
                    \hhconfig ; JM                \rangle,
\end{equation}
in fig.~\ref{f_gj}.
In the figure, $g_J^{(0)}$ ($g_J^{(1)}$) denotes the part of $g_J$
proportional to $u_0$ ($u_1$).  $g_J^{\rm (QQ)}$ is the matrix element
of $V_{\rm QQ}$ (eq.~(\ref{e_vqq})).  The range of fluctuation in
$g_J$ ($g_J^{\rm (QQ)}$) versus $J$ is 3.4~MeV (1.1~MeV).
In this respect, the former is stronger than the latter
by a factor of 3.0.

{}From the obtained enhancement factor of 1.5-3.0, $V_{\rm pn}$
does not seem unreasonably strong.  The difference of the strength
between $V_{\rm pn}$ of Semmes and Ragnarsson and
$\sim$10$\times V_{\rm QQ}$
of Hamamoto may originate in the $k \! \not= \! 2$
multipole interactions of the former force or in the difference of
configuration.

\subsection{The effect of the spin-dependent interaction} \label{s_spin}

Let us show that the spin-dependent term does not have strong
influence on the bands to be studied in this paper.
It has been known that
the proton-neutron interaction prefers spin-triplet state to singlet
one.  In deformed nuclei, it is realized as an empirical law called
the GM rule \cite{GM58}.
The GM splitting is the difference of energy between
states with $K_{>}$ and $K_{<}$
(see sect.~\ref{s_quart} for the definitions).
It is estimated
most simply with the difference in the diagonal matrix element of the
spin-dependent interaction between stretched and
antistretched unperturbed configurations.
For a product state of a proton and a neutron,
$
     |(l \mbox{\small $\frac{1}{2}$} ) j \Omega_{\rm p} \rangle_{\rm p}
     |(l \mbox{\small $\frac{1}{2}$} ) j \Omega_{\rm n} \rangle_{\rm n}
$,
the probabilities of spin singlet and triplet components are,
\begin{equation}
      {\rm Prob}. (S=1) = 1 - {\rm Prob}. (S=0) = \frac{3}{4} \; + \;
      \frac{\Omega_{\rm p} \Omega_{\rm n}}{4j^2}.
\end{equation}
In case of $^{124}$Cs, by using $j$ = $\frac{11}{2}$,
$|\Omega_{\rm p}|$ = $\frac{1}{2}$, and $|\Omega_{\rm n}|$ =
$\frac{7}{2}$, one finds only 3 \% difference of the probabilities
between configurations with $\Omega_{\rm p} \Omega_{\rm n} >0$ and
$<0$.  Therefore the GM splitting is very small.  The off-diagonal
effects are studied in sect.~\ref{s_res} (see fig.~\ref{f_u0}).

\section{The particle-triaxial-rotor model for odd-odd nuclei}
\label{s_model}

The hamiltonian of our model is expressed as,
\begin{equation} \label{e_ham}
        \hat{H} = \sum_{\kappa=1}^{3} \frac{\hbar^2}{2 {\cal I}_\kappa}
        (\hat{I}_{\kappa} - \hat{j}_{{\rm p} \kappa}
                          - \hat{j}_{{\rm n} \kappa})^2 +
	\hat{h}_{\rm Nilsson} + V_{\rm pn},
\end{equation}
where
\begin{equation}  \label{e_moi}
     {\cal I}_{\kappa} = \frac{4}{3}{\cal I}_0
     \sin^2(\gamma \mp \frac{2}{3} \pi \kappa )
     \;\;\; \mbox{for}
     \left\{
            \begin{array}{l}
                   \gamma \leq  0 \;\;\; \mbox{(irrotational flow)}, \\
                   \gamma > 0     \;\;\; \mbox{($\gamma$-reversed)}.
            \end{array}
     \right.
\end{equation}
$h_{\rm Nilsson}$ is a deformed single-particle potential,
which depends on $\gamma$ in such a way that
$\Delta \omega_{\kappa}$ $\propto$
$\cos (\gamma + \frac{2}{3} \pi \kappa )$.
$V_{\rm pn}$ is defined by eq.~(\ref{e_vpn}).
The subspace in which we diagonalize $\hat{H}$ is,
\begin{equation} \label{e_sp}
     \mbox{
     (1qp in $\pi$h$_{11/2}$) $\times$ (1qp in $\nu$h$_{11/2}$) $\times$
     ($|K_{\rm rotor}|$ = 0, 2, 4, 6).
}
\end{equation}
In eq.~(\ref{e_sp}), ``h$_{11/2}$'' stands for those Nilsson orbitals
which are continuously transformed to the h$_{11/2}$ spherical
orbitals as $\epsilon_2 \rightarrow 0$.  Orbitals with
$\Omega$=$\frac{11}{2}$ are not included.

The first term in the right-hand side of eq.~(\ref{e_ham}) is divided
into three parts in the usual way\cite{RS80}, {\it i.e.}, the
strong-coupling rotational energy, the Coriolis interaction, and
the recoil term.  In sect.~\ref{s_sys}, we use a Coriolis attenuation
factor $\rho$ of 0.7, which should be multiplied to all the
off-diagonal matrix elements of the Coriolis interaction.
Concerning the recoil term, operators
$\hat{\mbox{\boldmath $j$}}_{\rm p}^2$ and
$\hat{\mbox{\boldmath $j$}}_{\rm n}^2$
are treated as one-body operators
to avoid a double counting
of a rotor's contribution\cite{BLM87}.

The distinction between negative and positive values of $\gamma$ in
the definition (\ref{e_moi}) is introduced
just for the sake of convenience\cite{HM83}.  As for the
$\gamma$-dependence of the moment of inertia ${\cal I}_{\kappa}$, that
of irrotational flow is usually assumed.  Sometimes the
$\gamma$-reversed moment of inertia is also employed, which are
obtained by exchanging ${\cal I}_x$ and ${\cal I}_y$ by hand.  The
definition (\ref{e_moi}) is adopted so that the distinction between
the irrotational-flow moment of inertia and the $\gamma$-reversed one
can be made simply by means of the sign of $\gamma$.  Incidentally, in
this paper, a ``negative-$\gamma$ (positive-$\gamma$) rotor'' means a
rotor having an irrotational-flow ($\gamma$-reversed) moment of
inertia.

We employ the triaxial rotor model not because we believe in
rigid triaxial shapes but because it is a very convenient
model which can emulate the effects of fluctuation and vibration of
$\gamma$ around an axial shape.
Indeed, a triaxial rotor and a $\gamma$-vibrational axial rotor are
indistinguishable
to the first order in $\gamma$.  For more explanations of their
practical equivalence, see ref.~\cite{TO89} and references therein.

The parameters of the Nilsson potential are taken from table 1 of
ref.\cite{BR85}.  The pairing-active space is restricted to
$(15Z)^{1/2}$ proton and $(15N)^{1/2}$ neutron orbital pairs below and
above each fermi level.  The interaction strengths are obtained
by multiplying 0.95 to the standard strengths\cite{NT69} in order to
take into account the blocking effect.  The stretched coordinates are
used.

More explanations of the model are given in refs.\cite{SR90,SR92}.
Ref.~\cite{RS88} is about a simpler version of the model for 1qp bands
of odd-$A$ nuclei, originally introduced in ref.~\cite{MSD74},
and can be of some help to understand the present model.

The size of quadrupole deformation $\epsilon_2$ is taken from
ref.~\cite{CLJ92} for $^{120}$Cs and from ref.~\cite{KFH92} for
$^{124}$Cs.  A Nilsson-Strutinski calculation results in the same
$\epsilon_2$ for $^{124}$Cs\cite{Sh93}.  For other nuclei, it is
determined by averaging over neighboring even-even nuclei the values
of $\beta$ derived from B(E2)$\uparrow$\cite{RM87} .  When no data are
available, extrapolations are done.  Adopted values are given in the
second column of table \ref{t_parm}.

For the triaxiality parameter $\gamma$, we tried determining it by
equating $E(2^{+}_{\gamma})/E(2^{+}_{\rm gr})$ of a triaxial
rotor\cite{DF58} to the average of experimental values over
neighboring even-even nuclei\cite{Sa84}.  For $^{124}$Cs, we have
obtained $|\gamma| = 23^{\circ}$, which turns out to be a very
appropriate value concerning signature splitting.
Considering that this prescription to determine $\gamma$
is also successful for odd-$A$ nuclei\cite{Me75},
we adopt it for other nuclei, too.
The third column of table \ref{t_parm} presents the resulting values.

As a parameter to specify the moment of inertia, we use
$\ectp$, which stands for the smallest excitation energy of the bare
rotor, instead of ${\cal I}_0$.  In determining this parameter one has
to keep in mind that the moment of inertia of a real nucleus is not a
constant but usually increases versus spin.
As we are interested in signature inversions, which take place at
$I < I_{\rm inv} \cong 16$ (for $^{124}$Cs),
we determine $\ectp$ so as to fit $\Delta E/ \Delta I$ at
$I=I_{\rm inv}$, where $E$ is the eigenvalue of $\hat{H}$.  The
practical procedure of the fitting is as follows: For $^{120-128}$Cs
and $^{124-130}$La, we adjust $\ectp$ for each nucleus so that the
calculated $E(I=17)-E(I=15)$ agrees with the experimental energy
difference.  For $^{132}$La, $^{134}$Pr, and $^{136}$Pm,
$E(I=13)-E(I=11)$ is used for the fitting because no levels have been
observed for $I \geq 15$ for these nuclei.  The values in the fourth
column of table \ref{t_parm} are determined for $\rho=0.7$ and used in
section \ref{s_sys}.

\section{The results of calculations} \label{s_res}

In this section, we employ
residual pn interactions and $\gamma$-deformations and
examine the resulting
signature splittings and
electro-magnetic transition amplitudes.

We choose to define the signature splitting as follows: For each even
value of $I$, we interpolate the odd-$I$ sequence of the spectrum to
the even value of $I$ and
subtract the interpolated energy from the
even-$I$ energy level
\footnote{
For each even spin $I_0$, we choose the nearest $n$ odd spins to
$I_0$ and construct a polynomial in $I$ of degree
$n-1$ which passes through the $n$ levels for the $n$ spins (the
Lagrange interpolation formula).  It has been confirmed that the
result is practically the same for $n$=4, 6, and 8. We have observed
no artificial ripples due to our specific choice of the interpolation
method. We adopt $n$=6 in this paper.
}.
Negative values of the splitting mean signature inversion.
In this way we can treat the signature splitting
independent of the slope of the spectrum,
$\Delta E/ \Delta I$ ($\Delta I$=2),
and concentrate ourselves on the former quantity.
Indeed, the signature splittings
contain all the significant information of the spectra calculated in
subsect.~\ref{s_sys} since we use a parameter to fit
$\Delta E$ ($\Delta I$=2).

\subsection{Effects of pn interaction and $\gamma$-deformation}
\label{s_eff}

In this subsection, we try various combinations of two factors,
{\it i.e.}, pn force and triaxiality.
A nucleus $^{124}$Cs seems most appropriate
to this intensive study, since the experimental spin-assignments for
the nucleus are very reliable\cite{KFH92}.  $\ectp$ is fixed at
$0.15$~MeV for $^{120}$Cs and $0.20$~MeV for $^{124}$Cs.  The Coriolis
attenuation is not introduced ({\it i.e.}, $\rho$=1)
\footnote{
Although $\ectp$ is fixed, obtained spectrum is satisfactory for the
calculations given in fig.~\ref{f_u0}:
$\Delta E$ $\equiv$ $E(I=17)-E(I=15)$ is 90-100\%
of the experimental values.  For the calculations given in
figs.~\ref{f_nvpn} and \ref{f_wvpn}, $\Delta E$ is rather
small for $\gamma \not= 0^{\circ}$: It amounts to
$\sim$100\%, 70-80\%, and 60-70\% of the experimental value for
$\gamma$= $0^{\circ}$, $-23^{\circ}$, and $+23^{\circ}$, respectively.
To see the influence of such discrepancy, we have also done
calculations adjusting $\ectp$ for each value of
$\gamma$ to fit $\Delta E$.
The result supports the conclusion of this subsection,
provided $\rho$ $\sim$ $0.7$ (see sect.~\ref{s_sys}).
}.

First, we try changing $u_0$ while keeping axial symmetry. The value
of $u_1$ is fixed at $-0.8$~MeV.  The results for $^{120}$Cs and
$^{124}$Cs are shown in fig.~\ref{f_u0}.  The solid circles connected
with dashed curve designate the experimental signature splitting.  The
dots connected with solid curves are the calculated points for which
different values of $u_0$ ($0, -1.8, -3.6, -5.4, -7.2, -9.0$, and
$-10.8$ [MeV]) are used.  For $^{120}$Cs, $I_{\rm inv}$ coincides with
the experimental value when $u_0=-7.2$~MeV.  The slope is, however,
twice as small as the experimental trend.  (This is the same strength
as proposed by Semmes and Ragnarsson.  Our results seem in agreement
with fig.~7d of ref.~\cite{CLJ92}.)  For $^{124}$Cs, $I_{\rm inv}$ as
well as the slope are reproduced using a different strength
$u_0=-5.4$~MeV.  It is not preferable that the interaction strength
changes so widely between neighboring nuclei.

While all the solid curves in the figure are calculated with
$u_1$=$-0.8$~MeV, the plus marks connected with a dot-dash curve for
each nucleus are obtained without any pn interactions ($u_0=u_1=0$).
By comparing the curve with the top solid curve,
which is calculated with $u_0=0$ and $u_1=-0.8$~MeV,
one can see that a spin-dependent
interaction favoring $S$=1
enhances the normal-sign signature splitting, opposite to
the signature-inverting effect of spin-independent attractive
interactions.

In fig.~\ref{f_nvpn}, we tried introducing $\gamma$-deformations while
turning off the pn interaction ($u_0=u_1=0$).  With an axial rotor
($0^{\circ}$),
the inversion does not occur.  A negative-$\gamma$ rotor
($-23^{\circ}$) does not
change the result very much.  (See eq.~(\ref{e_moi}) for the
distinction between ``negative-$\gamma$'' and ``positive-$\gamma$''
rotors.)
For positive-$\gamma$ rotors ($+18^{\circ},+23^{\circ}$),
the signature splitting is sensitive to
the value of $\gamma$.  With $\gamma=+18^{\circ}$, $I_{\rm inv}$ as
well as the slope are reproduced.  As remarked in the introduction,
however, it is unlikely that real nucleus behaves like a
positive-$\gamma$ rotor.

In fig.~\ref{f_wvpn}, we show the results of calculations in which
$\gamma$-deformations as well as $V_{\rm pn}$ of Semmes and Ragnarsson
($u_0=-7.2$~MeV, $u_1=-0.8$~MeV) are taken into account.  With a
positive-$\gamma$ rotor ($+23^{\circ}$),
the effect to invert the signature splitting
is too strong, which is a natural result since both factors cooperate
to cause the inversion.
With a negative-$\gamma$ rotor ($-23^{\circ}$),
however, the
agreement with the experiment can be very good.

We imagine that a negative-$\gamma$ rotor plays the following role: At
low spins, the fluctuation in the orientation of the rotation axis is
so large that neither negative-$\gamma$ nor positive-$\gamma$ rotation
is realized. Consequently,
the static mean-field effect (such as considered
in the cranking model for negative-$\gamma$ rotation)
on the signature dependence of single-particle energy is
small\footnote{
At {\em very} low spins,
the angular-momentum-coupling scheme between
a particle and a rotor is not so simple as in this
argument\cite{Ha87,IA87}.
It seems related to the discussion in sect.~\ref{s_sign}.
It may be the reason why,
in our calculations,
negative-$\gamma$ rotors as well as positive-$\gamma$
ones seem to enhance the inversions.
}.
At high spins, the rotation axis is confined in the vicinity of the
intermediate-length axis because the excitation energy for wobbling
motion is proportional to $I$\cite{BM75}.  This rotation corresponds
to a negative-$\gamma$ rotation of the cranking model and contributes
to the normal-sign signature splitting.
Therefore, a negative-$\gamma$ rotor is expected to promote
the restoration of the normal-sign signature splitting at high spins
without hindering the signature inversion at low spins.

We now show that the combination of $V_{\rm pn}$ and a
negative-$\gamma$ rotor is not only physically most reasonable but
also supported by the experimental
B(M1;$I \rightarrow I-1$)/B(E2;$I \rightarrow I-2$) ratio.
In fig.~\ref{f_em}, this ratio is plotted as a function of $I$.
Calculations are done for three cases, {\it i.e.}, (1) when only a pn
interaction is taken into account (dotted lines, corresponding to the
curve using $u_0=-5.4$~MeV in the right-hand portion of
fig.~\ref{f_u0}), (2) when only a positive-$\gamma$ rotor is employed
(dashed lines, corresponding to the curve calculated with
$\gamma=+18^{\circ}$ in fig.~\ref{f_nvpn}),
and (3) when the pn interaction $V_{\rm pn}$
is combined with a negative-$\gamma$ rotor (solid lines,
corresponding to the curve using $\gamma=-23^{\circ}$ in
fig.~\ref{f_wvpn}).  In the left-hand portion, where the ratios for
odd values of $I$ are plotted, the solid line reproduces best the
decrease of the ratio versus $I$ (except for $I$=19).  In the
right-hand portion, which shows the ratios for even $I$, the sudden
drop in the ratio is reproduced only by the solid line.  (The drop is
attributed to the B(M1) value while B(E2) value increases
continuously.)

\subsection{A systematic calculation for $A \sim 130$ nuclei}
\label{s_sys}

We have seen that the combination of a pn interaction and a
negative-$\gamma$ rotor makes the most successful model to reproduce
signature-dependent quantities.  In this subsection we present
signature splittings calculated by this method for some Cs and La
isotopes and $N$=75 isotones.

Employed values for the parameters ($\epsilon_2$, $\gamma$, and $\ectp$)
are given in table \ref{t_parm}.
We use $V_{\rm pn}$ of Semmes and Ragnarsson
($u_0=-7.2$~MeV, $u_1=-0.8$~MeV).
The Coriolis attenuation factor of $\rho$=0.7 is assumed for all the
nuclei, the necessity of which is explained soon.

As explained in sect.~\ref{s_model}, we
choose to determine $\ectp$ so that $\Delta E$ for
$\Delta I$=2 is reproduced around $I_{\rm inv}$.  In
fig.~\ref{f_spc}, resulting spectra for $^{120,124}$Cs are shown.  The
solid (dashed) curves connect smoothly the calculated favored
(unfavored) signature levels.  Experimental favored (unfavored)
signature levels are denoted by solid (open) circles. The agreements
with experiment are quite excellent for wide ranges of $I$ centering
on $I=16$.  Equally excellent agreements are achieved for other
nuclei, too.

For the calculations in this section, we introduce the Coriolis
attenuation factor because the absolute values of the signature
splittings are too large when $\ectp$ is adjusted to reproduce the
$\Delta I$=2
transition energy.  The attenuation factor stands for many
ingredients
which are not taken into account in the particle-rotor model, some of
which may cooperate to weaken the Coriolis interaction.
In fig.~\ref{f_cori}, signature splittings calculated with various
values of $\rho$ are shown. We have adjusted $\ectp$ for each value of
$\rho$ to reproduce
$E(I$=17)$-E(I$=15)\footnote{
Because of this adjustment,
the ``rotational frequency'' $\omega_{\rm rot}=dE/dI$,
which is the strength of the Coriolis field in the cranking model,
is constant despite the change in $\rho$.
However, as intended, the Coriolis interaction
of our model is attenuated: It is proportional to a ratio
$\rho/\ectp$, which decreases as $\rho$ decreases ($\ectp$=0.28~MeV
for $\rho=1$ and $\ectp$=0.19~MeV for $\rho=0.6$).
}.
{}From this figure, one can see that $I_{\rm inv}$ as well as the slope
are reproduced excellently with $\rho=0.7$.  We use the same value of
$\rho$ for other nuclei, too.

The left-hand portion of fig.~\ref{f_cs} shows experimental signature
splittings of some Cs isotopes.  As for $^{120,124,126}$Cs, the
behaviors of signature splittings are rather similar to one another.
The curve for $^{128}$Cs seems to have the opposite sign to other
curves.  The right-hand portion gives the calculated splittings.  As
for $^{120}$Cs, the relatively large slope of the splitting is
reproduced.  Although the agreement is very good for $I \leq 12$ and
$I \geq 20$, there is a bump around $I \sim 15$ in the calculation and
hence $I_{\rm inv}$ is smaller than experiment by $\Delta I = -1.8$.
Concerning $^{124}$Cs, the agreement is quite satisfactory.
For $^{126}$Cs,
the calculated $I_{\rm inv}$ is larger than the experimental one
by $\Delta I = 2.0$.
The abrupt reversion of the sign of the splitting
between $A=126$ and $A=128$ is
not reproduced by our calculation.
The agreements of $I_{\rm inv}$
between the experiments and the calculations would become remarkably
excellent if the experimental spin-assignments could be changed by
$-2$, $+2$, and $+1$ for $^{120,126,128}$Cs, respectively.

In fig.~\ref{f_la}, some La isotopes are studied.
One can see a common feature
between the experiments and the calculations
that $I_{\rm inv}$ increases as $A$ increases.
However, while in experiment
signature splittings of these isotopes seem to have the opposite sign
to those of $^{120-126}$Cs, in our calculation
such opposite-sign behaviors between
$\Delta Z$=2 neighbors are not reproduced at all.
This systematic discrepancy
in the sign might signify that all the experimental spin-assignments
were incorrect by odd values of $\Delta I$ or it would indicate the
necessity of a more elaborate model.

The signature splittings of three $N$=75 isotones are shown in
fig.~\ref{f_n75}. In this case, the correspondence between the
experiments and the calculations is not very clear, partly because the
observed rotational sequences are very short.

\section{Summary} \label{s_sum}

In this paper, we have applied a particle-triaxial-rotor model to
odd-odd $A \sim 130$ nuclei to investigate the signature dependence of
the $\hhconfig$ bands.

In sect.~\ref{s_short}, we have explained two deficiencies of the
cranking model and questioned its applicability to signature-inversion
phenomena in odd-odd nuclei.  One is that there is not always clear
correspondence between the signature quantum number and the parity of
the total angular momentum.  The other is that single-particle
signatures are not conserved in multi-qp configurations.
We have also discussed about the signature quartette and
how its members are grouped into rotational bands.

In sect.~\ref{s_vpn}, we have compared
the zero-range residual interaction of Semmes and Ragnarsson
with a $\qq$ force having a standard strength
as for $\hhconfig$ configurations.
The former has turned out stronger than
the latter by a factor of 1.5-3.0 but not unreasonably strong.
The spin-dependent part of the force has also been discussed.

The model and the methods to determine the
parameters have been described in sect.~\ref{s_model}.

In sect.~\ref{s_eff}, we have tried various combinations of the
residual interactions and the rotors in order to reproduce the signature
splitting and the B(M1)/B(E2) ratio for $^{124}$Cs.
We have found that the best
result is obtained when the interaction of Semmes and Ragnarsson is
combined with a triaxial rotor with irrotational-flow moment of
inertia.  The residual interaction gives rise to the signature
inversion.  The triaxial rotor does not deteriorate but rather enhance
the inversion at low spins, while it promotes the restoration of the
normal-sign splitting at high spins.

In sect.~\ref{s_sys}, we have performed a systematic calculation of
the signature splittings for $^{120,124,126,128}$Cs,
$^{124,126,128,130}$La, $^{132}$La$_{75}$, $^{134}$Pr$_{75}$, and
$^{136}$Pm$_{75}$ following the prescription given in
sect.~\ref{s_eff}.
We need the Coriolis attenuation factor of 0.7 to
reproduce both the $\Delta I$=2 transition energy and the signature
splitting.
The agreement with experiment is good for Cs isotopes
except $^{128}$Cs.
For La isotopes and $^{128}$Cs,
the experimental splittings are of the opposite sign
to those of $^{120-126}$Cs.
Our calculation do not produce such opposite-sign behaviors
between $\Delta Z$=2 or $\Delta N$=2 nuclei at all.
For $N$=75 isotones,
the correspondence is not clear
between the experiments and the calculations.

As an indispensable step
to clarify the origin of the opposite-sign
signature splittings between experiment and theory,
we hope that more
reliable experimental spin-assignments are conducted
for nuclei other than $^{124}$Cs.

\vspace{\baselineskip}

\noindent
{\large {\bf Acknowledgements}}

The author thanks Drs.~I.~Ragnarsson and P.~Semmes for providing their
particle-triaxial-rotor model code for odd-odd nucleus and answering
some questions.  He is also grateful to Drs.~K.~Furuno and
T.~Komatsubara for useful information and discussion on the subject.
Most of the numerical calculations for this paper were performed on the
computer VAX 6440 of the Meson Science Laboratory, Faculty of Science,
University of Tokyo.  The author was supported in part by the
Grant-in-Aid for Scientific Research (no. 05740167) from the Ministry
of Education, Science and Culture of Japan.




\newpage

\noindent {\bf TABLE}

\newcounter{tabno}
\begin{list}
{TABLE \arabic{tabno}. }{\usecounter{tabno}
   \setlength{\labelwidth}{2cm}
   \setlength{\labelsep}{0.5mm}
   \setlength{\leftmargin}{15mm}
   \setlength{\rightmargin}{0mm}
   \setlength{\listparindent}{0mm}
   \setlength{\parsep}{0mm}
   \setlength{\itemsep}{0.5cm}
   \setlength{\topsep}{0.5cm}
}
\baselineskip=0.600cm
\item \label{t_parm}
    Adopted parameters for each nucleus.
    In sect.~\ref{s_eff}, we use $\ectp$ = $0.15$~MeV for $^{120}$Cs
    and $0.20$~MeV for $^{124}$Cs with $\rho$ = 1.
    In sect.~\ref{s_sys}, $\rho$ = 0.7 is employed.
\end{list}

\begin{table}[h]
\begin{tabular}{clcc}
\hline
                   & $\epsilon_2$ & $\gamma$     & $\ectp$ [MeV] \\
\hline
 $^{120}$Cs        & 0.25         &$-23^{\circ}$ & 0.159       \\
 $^{124}$Cs        & 0.22         &$-23^{\circ}$ & 0.206       \\
 $^{126}$Cs        & 0.244        &$-24^{\circ}$ & 0.217       \\
 $^{128}$Cs        & 0.210        &$-24^{\circ}$ & 0.182       \\
\hline
 $^{124}$La        & 0.31         &$-17^{\circ}$ & 0.182       \\
 $^{126}$La        & 0.288        &$-19^{\circ}$ & 0.191       \\
 $^{128}$La        & 0.268        &$-21^{\circ}$ & 0.200       \\
 $^{130}$La        & 0.246        &$-23^{\circ}$ & 0.210       \\
\hline
 $^{132}$La$_{75}$ & 0.217        &$-25^{\circ}$ & 0.179       \\
 $^{134}$Pr$_{75}$ & 0.225        &$-25^{\circ}$ & 0.206       \\
 $^{136}$Pm$_{75}$ & 0.225        &$-25^{\circ}$ & 0.224       \\
\hline
\end{tabular}
\end{table}


\newpage

\noindent {\bf FIGURE CAPTIONS}

\newcounter{figno}
\begin{list}
{Fig. \arabic{figno}. }{\usecounter{figno}
   \setlength{\labelwidth}{1.3cm}
   \setlength{\labelsep}{0.5mm}
   \setlength{\leftmargin}{8.5mm}
   \setlength{\rightmargin}{0mm}
   \setlength{\listparindent}{0mm}
   \setlength{\parsep}{0mm}
   \setlength{\itemsep}{0.5cm}
   \setlength{\topsep}{0.5cm}
}
\baselineskip=0.600cm
\item \label{f_q}
    Possible band structures of a signature quartette for
$(j_{\rm p})^1 (j_{\rm n})^1$-type configurations
(when $j_{\rm p}-j_{\rm n}$ is even).
RA (DA) means that the orbital is rotation- (deformation-) aligned.
$f$ ($u$) stands for the wavefunction for a
favored- (unfavored-) signature orbital.  In
the figure, we have omitted the subscripts p and n used in the text.
Instead, the first (second) symbol specifies the proton (neutron)
orbital.  For example, $fu$ in a box labeled by (RA,DA) means a
configuration in which a proton is in a favored-signature
rotation-aligned orbital and a neutron is in an unfavored-signature
deformation-aligned orbital.
\item \label{f_gj}
    Two-body matrix elements of the residual interaction proposed
by Semmes and Ragnarsson for $\hhconfig$ configurations ($g_J$).  The
contributions from its spin dependent and independent terms are also
shown ($g_J^{(1)}$ and $g_J^{(0)}$, respectively).  The matrix
elements of a $Q_{\rm p} \cdot Q_{\rm n}$ force with a standard
strength are included for the sake of comparison ($g_J^{(QQ)}$).
\item \label{f_u0}
    Signature splitting of the $\hhconfig$ band in $^{120}$Cs
(left-hand portion) and $^{124}$Cs (right-hand portion) calculated
with various values of $u_0$.  Axial rotors are used. See text for
explanations.
\item \label{f_nvpn}
    Signature splitting of the $\hhconfig$ band in $^{124}$Cs
calculated without $V_{\rm pn}$.  Solid and dot-dash curves are the
results using different values of $\gamma$. The experimental values
are expressed by solid circles connected with a dashed curve.
\item \label{f_wvpn}
    Same as in fig.~\ref{f_nvpn} but calculations are done using
$V_{\rm pn}$ of Semmes and Ragnarsson.
\item \label{f_em}
    B(M1;$I \rightarrow I-1$)/B(E2;$I \rightarrow I-2$) ratio of the
$\hhconfig$ band in $^{124}$Cs.  The ratios for odd (even) values of
$I$ are given in the left-hand (right-hand) portion.  The experimental
values are taken from fig.~2 (c) of ref.~\cite{KFH90} and expressed by
dots with error bars.  The dotted lines are calculated with
$\gamma=0^{\circ}$ and $u_0=-5.4$~MeV.  The dashed lines are
calculated with $\gamma=+18^{\circ}$ and $u_0=u_1=0$.  The solid lines
are calculated with $\gamma=-23^{\circ}$ and $u_0=-7.2$~MeV.
See text for explanations.
\item \label{f_spc}
    Rotational spectrum of the $\hhconfig$ band in $^{120}$Cs and
$^{124}$Cs.  The experimental energy levels of favored (unfavored)
signature are designated with solid (open) circles.  Solid (dashed)
curves pass through the calculated levels of favored (unfavored)
signature.  Used parameters are $\gamma=-23^{\circ}$, $u_0=-7.2$~MeV,
$u_1=-0.8$~MeV, and $\rho=0.7$ for both nuclei.  The parameter $\ectp$
has been chosen to fit experimental $E(I=17)-E(I=15)$.  A constant
energy is added to the calculated levels for each nucleus so that
$E(I=16)$ is equal to the experimental value.
\item \label{f_cori}
    Effects of the Coriolis attenuation factor $\rho$ on the
signature splitting of the
$(\pi{\rm h}_{11/2})^1$ $(\nu{\rm h}_{11/2})^1$ band in $^{124}$Cs.
Except for $\rho$, the same parameters as in fig.~\ref{f_spc}
are used.
Experimental splittings are represented by solid circles connected
with a dashed curve.  See text for explanations.
\item \label{f_cs}
    Experimental (left-hand side) and calculated (right-hand side)
signature splittings of $\hhconfig$ bands in some Cs isotopes.
Experimental data are taken from ref.~\cite{CLJ92} for $^{120}$Cs,
from ref.~\cite{KFH92} for $^{124,126}$Cs, and from ref.~\cite{PFL89}
for $^{128}$Cs.  No observed bands are ascribed to this configuration
for $^{122}$Cs\cite{XML90}.
\item \label{f_la}
    Experimental and calculated signature splittings of $\hhconfig$
bands in some La isotopes.
Experimental data are taken from
ref.~\cite{KFH92} for $^{124}$La, from ref.~\cite{NGB89} for
$^{126}$La, and from ref.~\cite{GHJ89} for $^{128,130}$La.
\item \label{f_n75}
    Experimental and calculated signature splittings of $\hhconfig$
bands in some $N$=75 isotones.  Experimental data are taken from
ref.~\cite{OER89} for $^{132}$La and from ref.~\cite{BHP87} for
$^{134}$Pr and $^{136}$Pm.
\end{list}


\begin{thebibliography}{99}
\bibitem{FM83} 
         S.\ Frauendorf and F.R.\ May,
	 Phys.\ Lett.\ {\bf B125} (1983), 245.
\bibitem{BFM84} 
         R.\ Bengtsson, H.\ Frisk, F.R.\ May, and J.A.\ Pinston,
	 Nucl.\ Phys.\ {\bf A415} (1984) 189.
\bibitem{OT88} 
         N.\ Onishi and N.\ Tajima,
	 Progr.\ Theor.\ Phys.\ {\bf 80} (1988) 130.
\bibitem{ALL76} 
         G.~Andersson, S.E.~Larsson, G.~Leander, P.~M\"{o}ller,
	 S.G.~Nilsson,
	 I.~Ragnarsson, S.~{\AA}berg, R.~Bengtsson, J.~Dudek,
	 B.~Nerlo-Pomorska, K.~Pomorski, and Z.~Szyma\'{n}ski,
	 Nucl.\ Phys.\ {\bf A268} (1976) 205.
\bibitem{KFH92} 
         T.~Komatsubara, K.~Furuno, T.~Hosoda, J.~Mukai,
	 T.~Hayakawa,
         T.~Morikawa, Y.~Iwata, N.~Kato, J.~Espino, J.~Gascon,
	 N.~Gj\/orup,
         G.B.~Hagemann, H.J.~Jensen, D.~Jerrestam, J.~Nyberg,
	 G.~Sletten,
         B.~Cederwall, and P.O.~Tj\/om,
	 Conf.  rapidly rotating nuclear
         structure, Tokyo, October 1992,
	 Nucl.Phys. {\bf A557} (1993) 419c.
\bibitem{HM83} 
         I.\ Hamamoto and B.R.\ Mottelson,
	 Phys.\ Lett.\ {\bf B132} (1983) 7.
\bibitem{BM75} 
         A.\ Bohr and B.R.\ Mottelson,
	 {\em Nuclear Structure}, vol.\ 2 \,
         (Benjamin, New York, 1975).
\bibitem{HH89} 
         J.~Helgesson and I.~Hamamoto,
	 Phys.Scripta. {\bf 40} (1989) 595.
\bibitem{Ta90} 
         N.~Tajima, Nucl. Phys. {\bf A520} (1990) 317c.
\bibitem{Ha86} 
 and sign.inv in 3qp band
         I.\ Hamamoto, Phys.\ Lett.\ {\bf B179} (1986) 327.
\bibitem{IS89} %
signature inversion using gamma-reversed m.o.i
         A.\ Ikeda and T.\ Shimano,
	 Phys.\ Rev.\ Lett.\ {\bf 63} (1989) 139.
\bibitem{SR90} 
         P.B.~Semmes and I.~Ragnarsson,
         Conf. high spin physics and gamma-soft
         nuclei, Pittsburg, September 1990
         (World Scientific, Singapore, 1991) p.500.
\bibitem{SR92} 
         P.B.~Semmes and I.~Ragnarsson,
	 Conf. future directions in nuclear
         physics with 4 $\pi$ detection systems of the new generation,
         Strasbourg, March 1991 (AIP, New York, 1992) p.566.
\bibitem{CLJ92} 
         B.~Cederwall, F.~Lid\'en, A.~Johnson, L.~Hildingsson, R.~Wyss,
         B.~Fant, S.~Juutinen, P.~Ahonen, S.~Mitarai,
	 J.~Mukai, J.~Nyberg, I.~Ragnarsson, and P.~Semmes,
	 Nucl. Phys. {\bf A542} (1992) 454.
\bibitem{Ma91} 
         M.~Matsuzaki, Phys. Lett. {\bf B269} (1991) 23.
\bibitem{Yo93} 
         N.~Yoshida, H.~Sagawa, and T.~Otsuka,
	 Nucl.Phys. {\bf A}, in press.
\bibitem{DF58} 
         A.S.~Davydov and G.F.~Filippov, Nucl.Phys. {\bf 8} (1958) 237.
\bibitem{Fr93} 
         S.~Frauendorf, Nucl.Phys. {\bf A557} (1993) 259c.
\bibitem{BP76} 
         J.P.~Boisson and R.~Piepenbring, Phys.Rep. {\bf 26} (1976) 99.
\bibitem{Se93} 
         P.~Semmes, private communication.
\bibitem{GM58} 
         C.J.~Gallagher and S.A.~Moszkowski, Phys.Rev. 111 (1958) 1282.
\bibitem{dF74} 
         A.~de Shalit and H.~Feshbach, Theoretical Nuclear Physics
	 (John Wiley and Sons, 1974), p.310.
\bibitem{dT63} 
         A.~de Shalit and I.~Talmi, Nuclear Shell Theory,
	 (Academic Press, New York, 1963), pp.213-214.
\bibitem{RS80} 
         P.~Ring and P.~Schuck, The nuclear many-body problem
         (Springer, New York, 1980)
\bibitem{BLM87} 
         L.~Bennour, J.~Libert, M.~Meyer, and P.~Quentin,
	 Nucl.\ Phys.\ {\bf A465} (1987) 35, see pp.~73-74.
\bibitem{TO89} 
         N.\ Tajima and N.\ Onishi, Nucl.\ Phys.\ {\bf A491} (1989) 179.
\bibitem{BR85} 
         T.~Bengtsson and I.~Ragnarsson, Nucl.Phys. {\bf A436} (1985) 14.
\bibitem{NT69} 
         S.G.~Nilsson, C.F.~Tsang, A.~Sobiczewski, Z.~Szyma\'{n}ski,
	 S.~Wycech, C.~Gustafson, I.~Lamm, P.~M\"{o}ller, and B.~Nilsson,
	 Nucl.Phys. {\bf A131} (1969) 1.
\bibitem{RS88} 
         I.~Ragnarsson and P.~Semmes, Hyp.Int. {\bf 43} (1988) 425.
\bibitem{MSD74} 
         J.~Meyer-ter-Vehn, F.S.~Stephens, and R.M.~Diamond,
	 Phys.\ Rev.\ Lett.\ {\bf 32} (1974) 1383.
\bibitem{Sh93} 
         Y.R.~Shimizu, private communication.
\bibitem{RM87} 
         S.~Raman, C.H.~Malarkey, W.T.~Milner, C.W.~Nestor, Jr., and
	 P.H.~Stelson, Atomic Data and Nucl. Data Tables, {\bf 36} (1987) 1.
\bibitem{Sa84} 
         M.~Sakai, Atomic Data and Nucl. Data tables {\bf 31} (1984) 399.
\bibitem{Me75} 
	 J.~Meyer-ter-Vehn, Nucl.\ Phys.\ {\bf A249} (1975) 111,141.
\bibitem{Ha87} 
         I.\ Hamamoto, Phys.\ Lett.\ {\bf B193} (1987) 399.
\bibitem{IA87} 
         A.~Ikeda and S.~{\AA}berg, Nucl.Phys. {\bf A480} (1988) 85.
\bibitem{KFH90} 
         T.~Komatsubara, K.~Furuno, T.~Hosoda, J.~Espino, J.~Gascon,
         G.B.~Hagemann, Y.~Iwata, D.~Jerrestam, N.~Kato, T.~Morikawa,
         J.~Nyberg, G.~Sletten, and P.O.~Tj\/om,
         Z. Phys. {\bf A335} (1990) 113.
\bibitem{PFL89} 
         E.S.~Paul, D.B.~Fossan, Y.~Liang, R.~Ma, and N.~Xu,
	 Phys.Rev. {\bf C40} (1989) 619.
\bibitem{XML90} 
         N.~Xu, Y.~Liang, R.~Ma, E.S.~Paul, D.B.~Fossan,
	 and H.M.~Latvakoski,
	 Phys.Rev. {\bf C41} (1990) 2681.
\bibitem{NGB89} 
         B.M.~Nyak\'{o}, J.~Gizon, D.~Barn\'{e}oud, A.~Gizon, M.~J\'{o}zsa,
	 W.~Klamra, F.A.~Beck, and J.C.~Merdinger,
	 Z.Phys. {\bf A332} (1989) 235.
\bibitem{GHJ89} 
         M.J.~Godfrey, Y.~He, I.~Jenkins, A.~Kirwan, P.J.~Nolan,
	 D.J.~Thornley, S.M.~Mullins, and R.~Wadsworth,
	 J.Phys. {\bf G15} (1989) 487.
\bibitem{OER89} 
         J.R.B.~Oliveira, L.G.R.~Emediato, M.A.~Rizzutto, R.V.~Ribas,
	 W.A.~Seale, M.N.~Rao, N.H.~Medina, S.~Botelho, and E.W.~Cybulska,
	 Phys.Rev. {\bf C39} (1989) 2250.
\bibitem{BHP87} 
         C.W.~Beausang, L.~Hildingsson, E.S.~Paul, W.F.~Piel, Jr., N.~Xu,
	 and D.B.~Fossan,
	 Phys.Rev. {\bf C36} (1987) 1810.
\end{thebibliography}
\end{document}